\documentclass[conference,letterpaper]{IEEEtran}
\IEEEoverridecommandlockouts

\usepackage{cite}
\usepackage{amsmath,amssymb,amsfonts}
\usepackage{graphicx}
\usepackage{textcomp}
\usepackage{xcolor}
\usepackage{booktabs}
\usepackage{multirow}
\usepackage{array}
\usepackage{url}

\title{
SERA-IDS: Structured Experience Retrieval-Augmented Intrusion Detection with Small Language Models
}

\author{
\IEEEauthorblockN{Md Shamimul Islam}
\IEEEauthorblockA{
\textit{Department of Computer Science}\\
\textit{Florida Polytechnic University}\\
Lakeland, FL, USA\\
mislam3051@floridapoly.edu
}
\and
\IEEEauthorblockN{Ayesha S. Dina}
\IEEEauthorblockA{
\textit{Department of Computer Science}\\
\textit{Florida Polytechnic University}\\
Lakeland, FL, USA\\
adina@floridapoly.edu
}
 \thanks{This work has been submitted to the IEEE for possible publication. 
Copyright may be transferred without notice, after which this version may no longer be accessible.}
}

\begin{document}
\maketitle

\begin{abstract}
Large language models (LLMs) offer a flexible approach to network intrusion detection, but direct classification of numerical flow records can be unreliable without traffic-specific decision boundaries. Retrieval-augmented generation (RAG) enables LLMs to leverage external experience; however, free-form textual experience requires the model to infer the numerical conditions that distinguish traffic classes. We propose \textbf{SERA-IDS}, a Structured Experience Retrieval-Augmented Intrusion Detection framework that learns decision knowledge from classification errors. A Small Language Model (SLM) analyzes misclassified flows and generates structured rules containing behavioral descriptions, feature conditions, class-confusion information, tool-derived confidence, and provenance. A confidence gate admits supported rules into the Experience Library. During testing, the library is frozen, and retrieved class-diverse rules are condition-matched to the query flow and provided as evidence to the decision SLM. The same library is evaluated with three small, locally deployable SLMs, namely Llama 3.1, Phi-4:14B, and Qwen2.5:7B, without fine-tuning or paid hosted inference. On the test sets, SERA-IDS improves macro F1 on NF-BoT-IoT from 11.66\%, 9.26\%, and 7.83\% to 86.46\%, 69.50\%, and 83.51\%, respectively, and on NF-ToN-IoT from 3.82\%, 6.17\%, and 1.94\% to 84.16\%, 87.66\%, and 83.37\%. These results show
that structured experience retrieval enables small, free, and
locally deployable SLMs to achieve strong intrusion-detection
performance, including performance exceeding that of the re-
ported existing work used in our comparison.
\end{abstract}



\section{Introduction}


Intrusion Detection Systems (IDSs) are essential for identifying malicious activities in modern computer networks. Traditional machine-learning and deep-learning approaches have achieved strong detection performance, but often require substantial labeled data and retraining when traffic characteristics change~\cite{denning1987,bace2001,khraisat2019}. Recently, LLMs have been explored for network security because of their ability to reason over contextual information~\cite{zhang2024,trafficgpt, moore2025}. 


However, deploying large language models can involve high computational
requirements, inference costs, and dependence on hosted services. In contrast, open-weight SLMs provide a practical alternative to paid hosted inference, with lower computational requirements and support for local
deployment. This makes them attractive for resource-conscious intrusion-detection systems. Language-model reasoning over numerical flow data remains challenging, especially for SLMs with limited domain knowledge. RAG supplies external knowledge at inference~\cite{rag2020}, but free-form experiences lack explicit numerical conditions, and a single retrieved experience may not resolve ambiguous flows. This motivates structured experiences with explicit feature-level conditions.

In this work, we propose SERA-IDS, a Structured Experience Retrieval-Augmented framework for SLM-based network intrusion detection. It converts training-time classification errors into structured rules containing feature conditions, target classes, confidence, and provenance, and a confidence gate admits only reliable rules into the active library. At inference, the system retrieves up to five class-diverse rules and combines semantic retrieval with condition verification for the final SLM decision, so the learned experience can be reused across SLMs without fine-tuning. We evaluate SERA-IDS on NF-BoT-IoT and NF-ToN-IoT with Llama 3.1, Phi-4:14B, and Qwen2.5:7B, comparing it with zero-shot classification and the earlier MA-IDS framework~\cite{maids2026}. Results show substantial Macro-F1 gains across models and datasets, and demonstrate that a library generated by one SLM supports classification with other SLMs.

The main contributions are a structured, confidence-gated experience-rule framework that learns from IDS classification errors, a diversity-aware top-five retrieval and condition-verification mechanism for SLM-based traffic classification, and a comparative evaluation across two NetFlow datasets and three SLMs. By combining structured external experience with SLM reasoning, this work provides a practical approach to improving network intrusion detection without model fine-tuning or continuous retraining.

\section{Background and Related Work}


\subsection{Intrusion Detection for IoT Networks}
The foundations of network intrusion detection include misuse/signature and anomaly-oriented approaches~\cite{denning1987,bace2001}. Machine-learning and deep-learning methods are widely applied to network traffic, with persistent challenges including class imbalance, concept drift, false alarms, computational cost, and explainability~\cite{shone2018,vinayakumar2019,khraisat2019,mltaxonomy2022}. Flow-based benchmarks are particularly relevant because they capture behavioral and temporal traffic characteristics without requiring packet payload inspection. NF-BoT-IoT-v3 and NF-ToN-IoT-v3 provide large-scale NetFlow-based traffic with multiple attack classes and temporal flow attributes~\cite{nfiotdatasets,netflowtemporal,tonv3}, making them suitable for evaluating structured experience retrieval for SLM-based intrusion detection.

\subsection{LLMs and Agentic Security Analysis}
Recent advances in language modeling have introduced a different paradigm for security analysis. Transformer language models changed the scale of general-purpose language reasoning~\cite{vaswani2017,brown2020}. More recent open-weight SLMs, including Llama 3.1~\cite{llama3}, Qwen2.5~\cite{qwen25}, and Phi-4~\cite{phi4}, enable language inference at substantially smaller parameter scales than frontier-scale models~\cite{slmsurvey2025}, creating an opportunity for security systems that need local execution, controlled data handling, or predictable operating costs. Throughout this paper, SLM refers to compact, locally deployable open-weight models in the 7B--14B parameter range.

LLM-based network security research has begun to explore this direction. Zhang et al. showed that in-context examples can improve automatic intrusion detection~\cite{zhang2024}. TrafficGPT studied LLM-based encrypted traffic classification~\cite{trafficgpt}. IDS-Agent combines an LLM with tools, knowledge, and memory for explainable IoT intrusion detection~\cite{idsagent}. NetMoniAI adopts a multi-agent architecture that distributes monitoring and analysis across local agents while maintaining centralized coordination~\cite{netmoniai}. A recent survey frames LLMs as processors, detectors, explainers, and controllers within cognitive NIDS~\cite{yang2025}. While these studies demonstrate the usefulness of contextual information for security analysis, the effect of experience representation and retrieval size on SLM-based network-flow classification remains less explored.

\subsection{Retrieval-Augmented Reasoning}
RAG combines parametric language models with an external non-parametric memory~\cite{rag2020}. Dense retrieval and sentence embeddings allow semantically related records to be found even when their surface forms differ~\cite{sbert}. FAISS provides efficient similarity search over large vector collections~\cite{faiss}. In security, retrieval can preserve prior evidence, incident knowledge, or analyst decisions without modifying the base model. The key design choice is therefore not merely whether to retrieve, but what constitutes a useful retrieval unit.

Our approach treats an experience rule as a hybrid object: its descriptor is optimized for semantic retrieval, while its conditions are retained for explicit feature-level interpretation. This is different from retrieving a raw historical flow or a prose-only explanation. It also differs from fine-tuning because the decision model parameters remain unchanged. This work therefore examines the effect of structured external memory on SLM-based flow classification.

\section{Methodology}
\subsection{System Overview}
\begin{figure*}[t]
    \centering
    \vspace{-2mm}
    \includegraphics[width=0.92\textwidth]{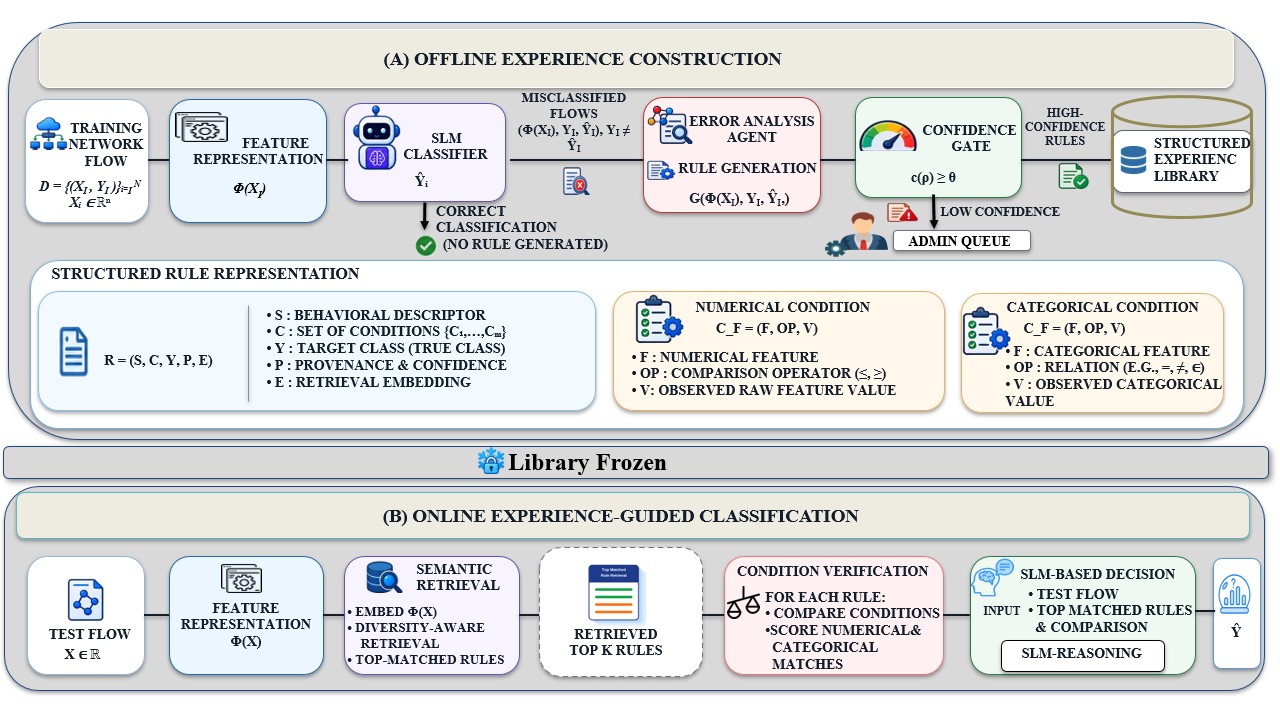}
    \vspace{-2mm}
    \caption{Overview of the proposed SERA-IDS framework. (a) Offline rule construction from misclassified training flows, including error analysis, confidence-based admission, and structured rule storage. (b) Online classification using feature representation, diversity-aware top-five retrieval, condition verification, and SLM-based decision reasoning.}
    \label{fig:overview}
    \vspace{-3mm}
\end{figure*}

The proposed framework separates experience construction from final classification. During construction, a Llama 3.1 SLM classifier processes training flows and, only for misclassified flows, an Error Analysis Agent (EAA) converts the observed confusion into a structured experience rule. A confidence gate determines which candidate rules become active retrieval entries. During inference, the resulting library is frozen; each held-out flow is embedded, class-diverse rules are retrieved, their conditions are verified, and the selected SLM produces the final class without parameter updates. Figure~\ref{fig:overview} summarizes the complete workflow.

Let $x$ denote a flow and $\mathcal{L}$ the active experience library. The retrieval and final decision operations are defined in Section~\ref{sec:retrieval-verify}.

\subsection{Structured Experience Representation}\label{sec:structured-repr}
Each learned experience is represented as
\begin{equation}
r=(s,C,y,p,e),
\label{eq:rule}
\end{equation}
where $s$ is a concise behavioral descriptor, $C$ is the structured condition body, $y$ is the target class, $p$ is provenance and confidence metadata, and $e$ is the retrieval embedding. The condition body contains numerical and categorical constraints. Unlike a prose-only experience, the rule preserves the measurable attributes that characterize the decision region.

For a numerical feature $f$, a condition is represented as
\begin{equation}
c_f=(f,op,v),
\label{eq:numcondition}
\end{equation}

where $v$ is the observed raw feature value associated with the rule and $op$ specifies the directional relation between the feature and the condition threshold. In the reported implementation, the operator is determined from the feature-wise empirical distribution of the observed value. Specifically, the observed value is mapped to its normalized position in the feature distribution; values below the feature-wise median position induce $op=\leq$, whereas values at or above the median position induce $op=\geq$. The raw value $v$ is retained as the condition threshold. Thus, the feature-level distribution determines the direction of the condition while the specific misclassified flow supplies the threshold value. Categorical conditions similarly retain explicit observed values and comparison relations. During inference, these conditions are evaluated against the corresponding feature values of the current flow to provide explicit numerical and categorical evidence alongside semantic rule retrieval.

The provenance record includes the target class, the class involved in the original error, support information, confidence, construction round, and source chunk. The confused-class information is especially useful for error-derived learning: the rule is intended to capture evidence that distinguishes two classes that the classifier actually confused, rather than merely restating a generic description of an attack family.

\subsection{Error-Driven Experience Construction}
The condition bodies and provenance fields are populated through an error-driven construction process. For a training flow $x$ with true label $y$ and classifier output $\tilde{y}$, rule construction is triggered only when $\tilde{y}\neq y$. The Error Analysis Agent (EAA) receives the flow context and both labels, analyzes the observed confusion, and produces a candidate structured rule:
\begin{equation}
 \rho=G(\phi(x),y,\tilde{y}), \qquad \tilde{y}\neq y,
\label{eq:rulegen}
\end{equation}
where $\phi(x)$ denotes the structured feature representation of flow $x$, and $\rho$ denotes a candidate rule before admission to the active library. The EAA generates the behavioral description and condition information, while the fixed-uniform condition feature configuration described in Section~\ref{sec:datasets}  is used consistently across generated rules. This common schema makes the resulting conditions directly comparable during retrieval and verification.


\subsection{Confidence-Gated Rule Admission}
A candidate rule is admitted only when sufficient external evidence supports its reliability. The EAA passes the flow and its generated description to a lightweight ToolSuite that evaluates $N$ complementary signals: an abuse-reputation-style score, a malicious/unknown/benign assessment, a threat-pulse-style activity score, and similarity between the generated technique description and reference attack-technique descriptions. Each signal $s_i\in[0,1]$ is normalized to a common scale, and the admission confidence is computed as
\begin{equation}
c(\rho)=\left(\frac{1}{N}\sum_{i=1}^{N}s_i\right)\left(\frac{1}{N}\sum_{i=1}^{N}\mathbf{1}[s_i\geq\tau]\right).
\label{eq:confidence}
\end{equation}
Thus, confidence reflects both the average strength of the available evidence and the consistency among the $N$ signals, where $\tau$ is the per-signal agreement threshold. An admission threshold $\theta$ is applied: candidates with $c(\rho)\geq\theta$ are admitted to the active experience library, whereas lower-confidence candidates are placed in an administrative queue and excluded from retrieval. In the reported experiments, $N= 4$ and $\tau=0.5$. The $N$ signals are evaluated against an offline synthetic testbed rather than live Internet services, providing a reproducible proxy for external threat corroboration rather than a real-time intelligence feed. Accepted rules are appended individually without deduplication or merging, preserving an auditable record of error-derived experience.

\subsection{Diversity-Aware Retrieval and Condition Verification}\label{sec:retrieval-verify}
For each flow, its $d$-dimensional embedding is queried against the active FAISS index~\cite{faiss}. A candidate pool of $P$ nearest rules is first retrieved. For each represented class $y$, the highest-similarity rule is retained, and the $K$ classes with the highest corresponding similarities form the class-diverse retrieval set:
\begin{equation}
r_y^*(x)=
\arg\max_{r\in\mathcal{C}_P(x)\cap\mathcal{L}_y}
\operatorname{sim}(e_x,e_r),
\label{eq:bestclassrule}
\end{equation}

\vspace*{4pt}
\begin{equation}
\mathcal{R}_K(x)=
\operatorname{TopK}_{y}
\left\{
\left(r_y^*(x),
\operatorname{sim}(e_x,e_{r_y^*(x)})\right)
\right\}.
\label{eq:retrieval}
\end{equation}

Here, $\mathcal{C}_P(x)$ denotes the $P$ nearest active rules to $x$, $\mathcal{L}_y$ contains active rules of class $y$, and $\operatorname{TopK}_y$ selects the $K$ classes with the highest best-rule similarity.


Semantic similarity is complemented by structured condition comparison.
Numerical conditions retain the raw feature value and the comparison
operator derived from its feature-wise normalized position, while
categorical conditions preserve explicit observations such as
destination port, protocol, and TCP flags. The query flow is compared with the retrieved conditions to produce feature-level evidence, denoted by $V(x,\mathcal{R}_K(x))$.

The final decision is based on the flow, retrieved rules, and condition evidence:
\begin{equation}
\hat{y}=A_m\!\left(x\;\Vert\;\mathcal{R}_K(x)\;\Vert\;V(x,\mathcal{R}_K(x))\right),
\label{eq:decision}
\end{equation}
where $A_m$ denotes the selected SLM. The condition evidence provides contextual support rather than a deterministic vote, and the SLM produces the final class prediction.

\section{Experimental Results}

\subsection{Experimental Platform and Protocol}
All experiments were conducted on Google Colab using an NVIDIA T4 GPU. The same hardware was used for all zero-shot, MA-IDS, and SERA-IDS evaluations to ensure consistent latency comparisons. For each dataset, 50,000 construction flows and 20,000 disjoint evaluation flows are drawn by uniform random sampling with equal class quotas, following MA-IDS~\cite{maids2026}; classes with fewer available flows, such as Theft and Ransomware, contribute all of their flows. Llama 3.1 generates candidate rules only for misclassified training flows; the resulting library is then frozen, and test labels are used solely for metric computation. No new rules are added and no existing rule is modified during testing.

\subsection{Datasets and Network Flow Representation}\label{sec:datasets}
We evaluate NF-BoT-IoT-v3 and NF-ToN-IoT-v3~\cite{nfiotdatasets,tonv3,netflowtemporal}. NF-BoT-IoT is evaluated using five classes: Benign, DDoS, DoS, Reconnaissance, and Theft. NF-ToN-IoT is evaluated using ten classes: Benign, Scanning, DDoS, Backdoor, DoS, Injection, Password, XSS, MITM, and Ransomware. Rare classes are retained in both multiclass tasks.


Each flow is represented using 14 numerical NetFlow features together
with categorical and protocol-related information. The numerical
features capture complementary traffic characteristics, including byte
and packet volumes, flow duration, bidirectional inter-arrival times,
bidirectional throughput, TCP-window statistics, longest-packet length,
and minimum/maximum TTL. Additional categorical context includes
destination port, protocol, L7 protocol, and TCP flag information.
Although the datasets provide 53 NetFlow attributes, this reduced
representation retains a compact set of complementary behavioral
signals for intrusion discrimination~\cite{nfiotdatasets,netflowtemporal}.
The same representation is used for both rule construction and
inference. During rule construction, a fixed-uniform feature
configuration is applied, such that every generated rule uses the same
predefined numerical and categorical feature sets rather than selecting
a different subset for each rule. For semantic retrieval, rule descriptors and query flows are embedded using all-MiniLM-L6-v2~\cite{sbert}, while the original numerical and categorical feature values are preserved for explicit condition verification.


\subsection{Experimental Configuration}
The reported experiments use a fixed embedding dimension $d=384$, retrieval candidate-pool size $P=20$, class-diverse retrieval width $K=5$, and confidence-admission threshold $\theta=0.80$, where $d$ is the semantic embedding dimension, $P$ the number of nearest active rules considered as the retrieval candidate pool, $K$ the maximum number of class-diverse rules retained per query, and $\theta$ the minimum confidence required for automatic admission into the active Experience Library. Candidates below $\theta$ are assigned to the administrative queue and excluded from active retrieval. These values are fixed across both datasets and all three SLMs.

\subsection{Decision Models and Baseline Comparisons}
The evaluation uses three open-weight SLMs that can be deployed locally: Llama 3.1 (8B parameters, Meta), Phi-4 (14B parameters, Microsoft), and Qwen2.5 (7B parameters, Alibaba)~\cite{llama3,phi4,qwen25}. Their local deployment avoids dependence on a hosted inference API during evaluation and permits the same external Experience Library to be tested across models. The frozen Experience Library is constructed by Llama 3.1 and then reused by all three decision models. No parameter fine-tuning or model updates are performed during evaluation, and all decision runs use temperature zero. The SLM produces the final classification from the flow, retrieved experience, and condition evidence while keeping the pretrained model parameters fixed.



We compare three configurations. Zero-Shot (ZS) provides the flow representation to the SLM without access to the learned experience library. The earlier MA-IDS configuration~\cite{maids2026} employs a free-form experience representation with single-rule retrieval, whereas SERA-IDS employs a structured experience representation with top-five retrieval. This comparison jointly varies rule representation and retrieval width, moving from a single free-form experience to five structured experiences with explicit conditions, and therefore evaluates the combined redesign rather than isolating either factor independently.

\subsection{Evaluation Metrics}
We report accuracy, macro recall, macro precision, and macro F1 as the primary classification metrics. Macro F1 is emphasized because both datasets contain multiple intrusion classes, including relatively rare classes, making a class-balanced summary useful for comparison across the multiclass setting. Average latency is reported separately in seconds per flow for the complete evaluation runs, capturing the end-to-end inference cost of each configuration.

\subsection{Structured Experience Library Construction}
\begin{figure}[t]
    \centering
    \includegraphics[width=\columnwidth]{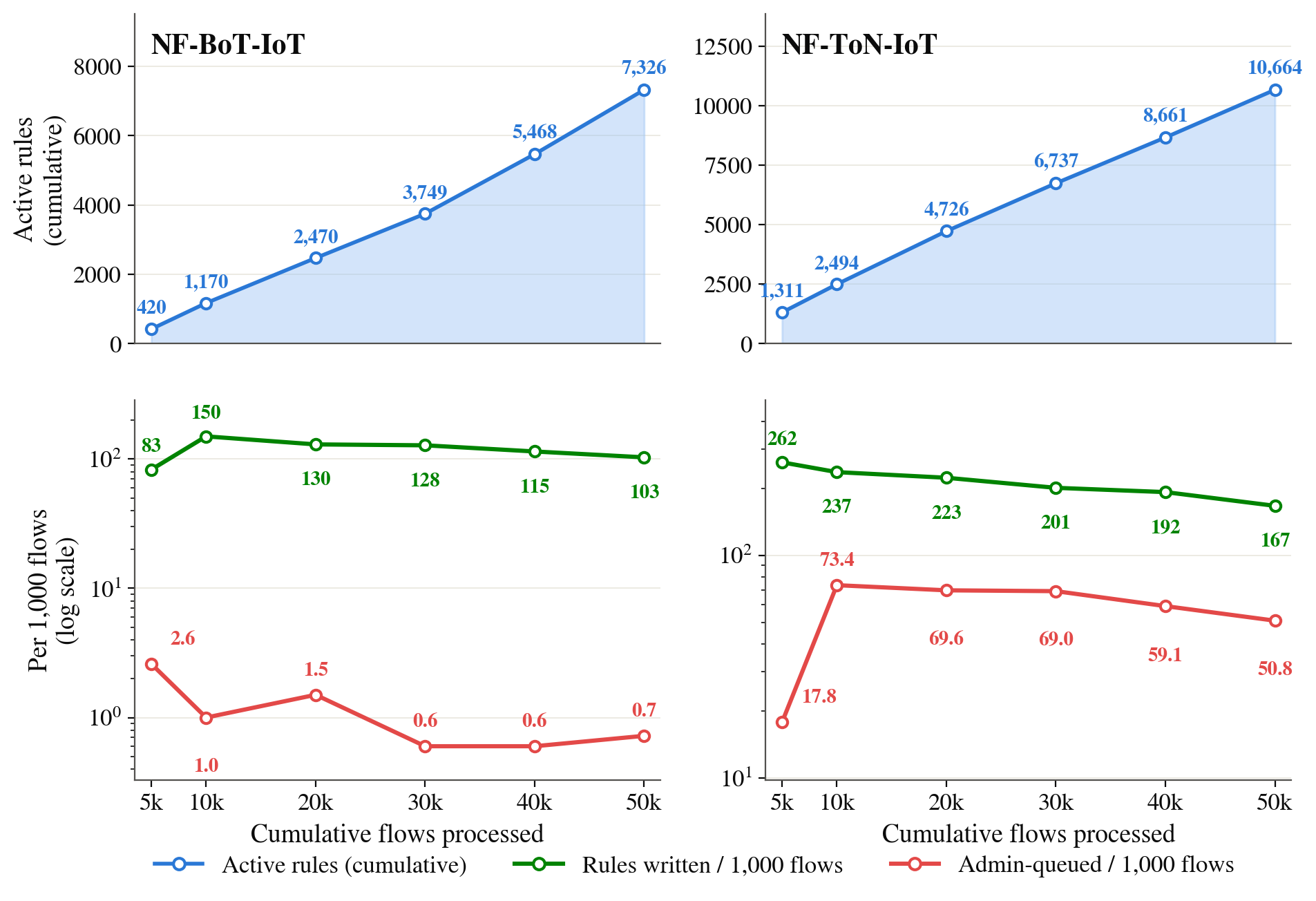}
    \caption{Experience-Library learning dynamics during construction. The upper panels show cumulative active rules, while the lower panels show rules written and administrative-queue rates per 1,000 processed flows for NF-BoT-IoT and NF-ToN-IoT.}
    \label{fig:library_growth}
\end{figure}

Figure~\ref{fig:library_growth} summarizes the evolution of the Experience Library during construction, including cumulative rule growth and rule-creation/admission dynamics without a separate construction-statistics table.

The construction process is error-driven: correctly classified flows do not generate new rules, whereas misclassified flows may yield high-confidence rules for the active library or lower-confidence candidates for administrative review. By the end of construction, the active libraries contained 7,326 rules for NF-BoT-IoT and 10,664 rules for NF-ToN-IoT, while 56 and 3,043 lower-confidence candidates, respectively, were assigned to the administrative queue. As shown in Fig.~\ref{fig:library_growth}, rule-generation rates generally decline at later construction stages, suggesting increasing coverage of recurring classification errors by the accumulated experience. The substantially larger administrative queue for NF-ToN-IoT also indicates that more candidate rules failed the confidence-admission criterion.

\subsection{Performance Comparison: Zero-Shot, MA-IDS, and SERA-IDS}
Table~\ref{tab:main} summarizes the principal test results. Zero-shot macro F1 is low for all three SLMs (1.94\% to 11.66\%), indicating that the models lack reliable decision boundaries for raw numerical flow features. SERA-IDS improves macro F1 for every model-dataset pair, from 11.66\%, 9.26\%, and 7.83\% to 86.46\%, 69.50\%, and 83.51\% on NF-BoT-IoT for Llama 3.1, Phi-4, and Qwen2.5, respectively, and by 80.34, 81.49, and 81.43 points on NF-ToN-IoT. Since each comparison holds the decision model fixed, these gains are associated with the structured Experience Library.

\begin{table}[t]
\centering
\caption{Performance comparison of three SLM-based intrusion detection settings, Zero-Shot (ZS), the earlier MA-IDS framework, and the proposed SERA-IDS framework, on NF-BoT-IoT and NF-ToN-IoT.}
\label{tab:main}
\scriptsize
\setlength{\tabcolsep}{3pt}
\renewcommand{\arraystretch}{0.92}
\begin{tabular*}{\columnwidth}{@{\extracolsep{\fill}}lrrrr@{}}
\toprule
\textbf{Model / Setting} &
\textbf{Acc.} & \textbf{Rec.} & \textbf{Prec.} & \textbf{F1} \\
\midrule
\multicolumn{5}{c}{\textit{NF-BoT-IoT}}\\[-2pt]
\cmidrule(lr){1-5}
Llama 3.1--ZS       & 20.77 & 14.20 & 13.50 & 11.66\\
Llama 3.1--MA-IDS   & 73.00 & 67.60 & \textbf{80.60} & 69.60\\
Llama 3.1--SERA-IDS & \textbf{84.66} & \textbf{94.78} & 79.48 & \textbf{86.46}\\
\cmidrule(lr){1-5}
Phi-4:14B--ZS       & 15.45 & 12.87 & 36.72 & 9.26\\
Phi-4:14B--MA-IDS   & 45.00 & 33.20 & 35.20 & 29.80\\
Phi-4:14B--SERA-IDS & \textbf{77.35} & \textbf{96.34} & \textbf{54.36} & \textbf{69.50}\\
\cmidrule(lr){1-5}
Qwen2.5:7B--ZS       & 24.31 & 19.95 & 8.20 & 7.83\\
Qwen2.5:7B--MA-IDS   & 69.00 & 63.60 & \textbf{80.00} & 63.40\\
Qwen2.5:7B--SERA-IDS & \textbf{81.77} & \textbf{95.14} & 74.41 & \textbf{83.51}\\
\midrule
\multicolumn{5}{c}{\textit{NF-ToN-IoT}}\\[-2pt]
\cmidrule(lr){1-5}
Llama 3.1--ZS       & 10.45 & 8.98 & 13.62 & 3.82\\
Llama 3.1--MA-IDS   & 67.00 & 60.70 & \textbf{80.70} & 64.30\\
Llama 3.1--SERA-IDS & \textbf{83.21} & \textbf{95.57} & 75.18 & \textbf{84.16}\\
\cmidrule(lr){1-5}
Phi-4:14B--ZS       & 14.46 & 12.34 & 21.38 & 6.17\\
Phi-4:14B--MA-IDS   & 12.00 & 17.40 & 50.70 & 15.30\\
Phi-4:14B--SERA-IDS & \textbf{87.13} & \textbf{97.14} & \textbf{79.87} & \textbf{87.66}\\
\cmidrule(lr){1-5}
Qwen2.5:7B--ZS       & 10.63 & 9.99 & 1.16 & 1.94\\
Qwen2.5:7B--MA-IDS   & 77.00 & 78.10 & \textbf{80.70} & 78.60\\
Qwen2.5:7B--SERA-IDS & \textbf{82.56} & \textbf{95.32} & 74.08 & \textbf{83.37}\\
\bottomrule
\end{tabular*}
\end{table}

\begin{figure}[t]
    \centering
    \includegraphics[width=\columnwidth]{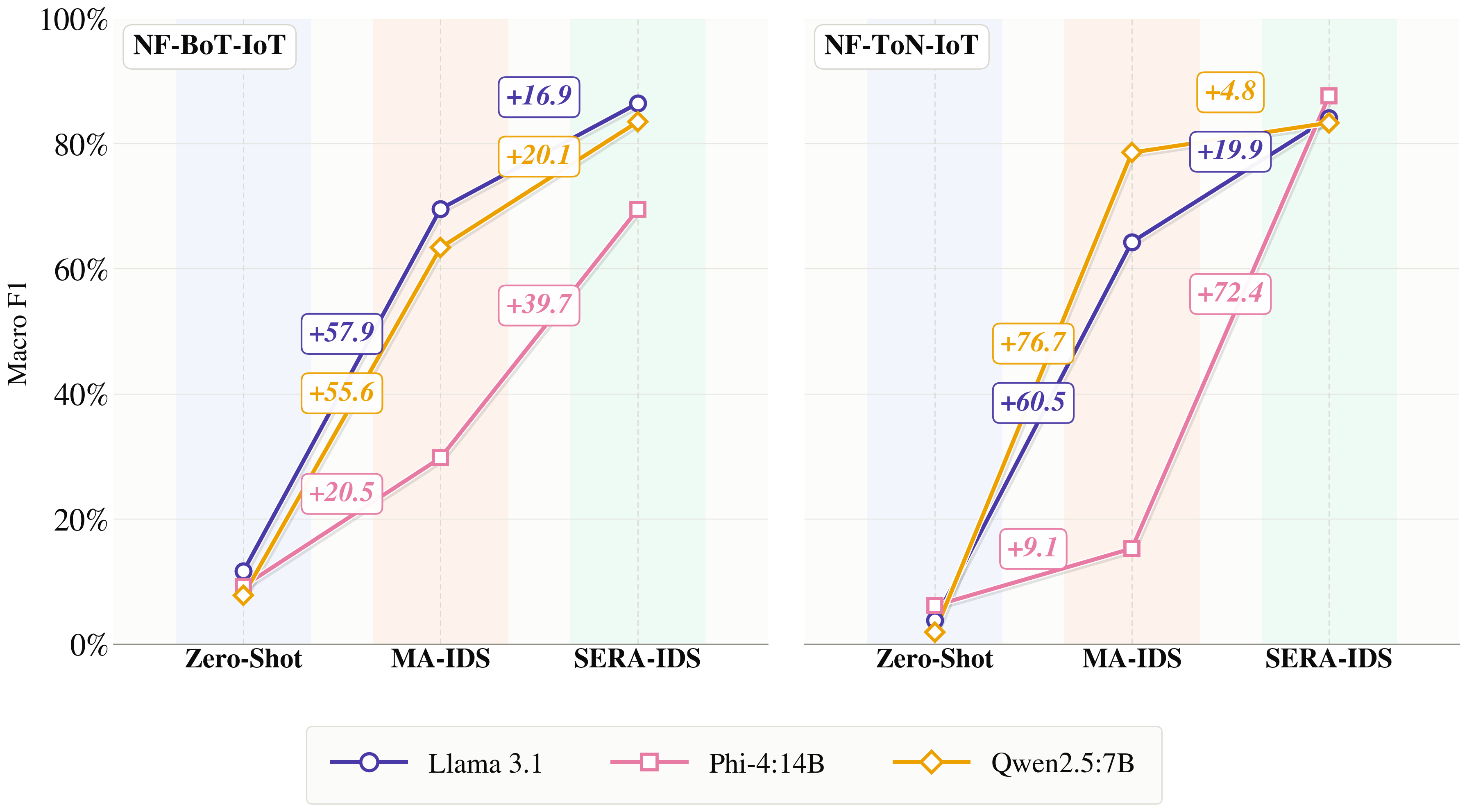}
    \caption{Macro-F1 progression from zero-shot classification to the earlier MA-IDS framework and SERA-IDS. Annotations indicate the corresponding Macro-F1 changes in percentage points.}
    \label{fig:f1_slope}
\end{figure}

As shown in Fig.~\ref{fig:f1_slope}, macro F1 rises monotonically from zero-shot through MA-IDS to SERA-IDS for every model on both datasets. Relative to MA-IDS, SERA-IDS adds 16.86, 39.70, and 20.11 points on NF-BoT-IoT and 19.86, 72.36, and 4.77 points on NF-ToN-IoT.

Under SERA-IDS, macro recall stays between 94.78\% and 97.14\% for all models and datasets, whereas macro precision varies from 54.36\% to 79.87\%. Compared with MA-IDS, SERA-IDS lowers macro precision for Llama 3.1 and Qwen2.5 by 1.12 to 6.62 points but raises their macro recall by 17.22 to 34.87 points. Since macro recall weights every class equally, this indicates broader coverage of rare attack classes at the cost of more inter-class confusion. One plausible cause is that five class-diverse rules expose the SLM to more candidate classes than a single experience, although this study does not isolate retrieval width from rule representation. This trade-off generally suits intrusion detection, where missed attacks typically cost more than false alarms an analyst can dismiss.

The benefit also depends on the model. Phi-4 reaches only 15.30\% macro F1 on NF-ToN-IoT with free-form MA-IDS experience but 87.66\% with SERA-IDS, the highest result in Table~\ref{tab:main}, suggesting that explicit feature conditions are easier for it to use than prose descriptions. On this dataset, Phi-4 also outperforms Llama 3.1, which constructed the library, indicating that the structured rules transfer across decision models.

\subsection{Runtime and Latency Analysis}
\begin{figure}[t]
    \centering
    \includegraphics[width=\columnwidth]{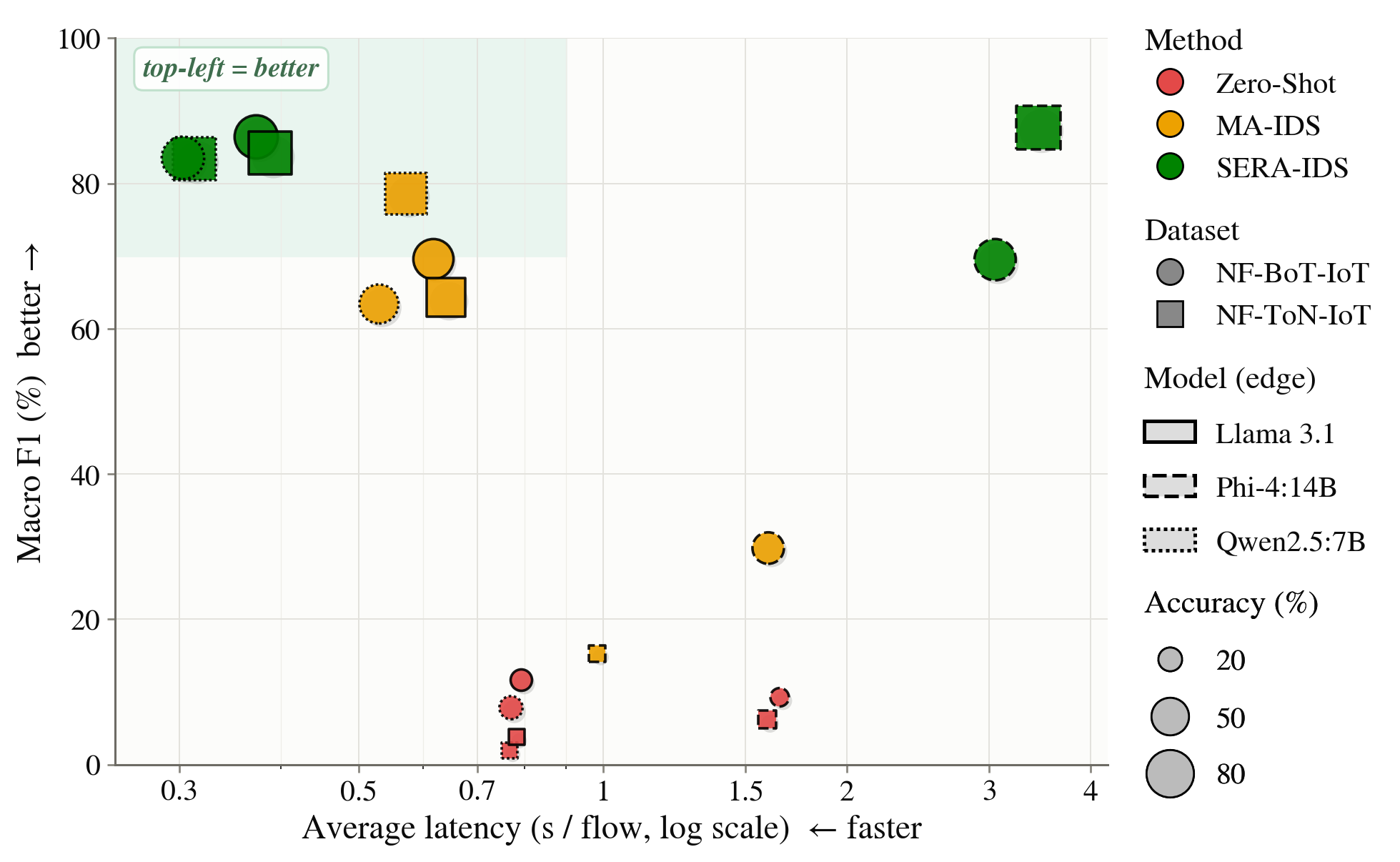}
    \caption{Latency--performance landscape across zero-shot, earlier MA-IDS, and SERA-IDS configurations. The x-axis shows average latency per flow on a logarithmic scale, the y-axis shows Macro-F1, and bubble size represents accuracy.}
    \label{fig:latency_bubble}
\end{figure}

Figure~\ref{fig:latency_bubble} shows the latency--performance trade-off. With SERA-IDS, Llama 3.1 and Qwen2.5 stay below 0.4~s/flow, whereas Phi-4 requires 3.05 to 3.45~s/flow, about an order of magnitude slower, without consistently better detection, since it achieves the best macro F1 on NF-ToN-IoT but the lowest on NF-BoT-IoT. In our setting, Llama 3.1 and Qwen2.5 therefore provide a more favorable balance between detection performance and inference cost. These are end-to-end local inference measurements rather than line-rate IDS throughput, suggesting that SERA-IDS may suit the analysis of flows pre-filtered by a lightweight detector better than packet-rate detection.


\section{Conclusion}
This paper presented SERA-IDS, a structured experience retrieval-augmented framework for SLM-based network intrusion detection. Llama 3.1 constructs an error-driven library of structured rules containing semantic descriptors, feature conditions, confidence, and provenance, while inference uses class-diverse top-five retrieval and condition verification. Across NF-BoT-IoT and NF-ToN-IoT, SERA-IDS substantially improves Macro-F1 over zero-shot classification for Llama 3.1, Phi-4:14B, and Qwen2.5:7B without fine-tuning, and the same Llama-generated library can be reused across all three SLMs.

The approach also extends our earlier GPT-4o-based MA-IDS study~\cite{maids2026}. On NF-ToN-IoT, Phi-4:14B with SERA-IDS achieves 87.66\% Macro-F1, exceeding the earlier GPT-4o result of 85.22\%, although the present evaluation uses ten classes compared with nine previously. These results show that structured external experience improves SLM-based network-flow classification without hosted inference. Future work will explore component ablations, unseen attacks, and library robustness.


\end{document}